\documentclass[aps,twocolumn]{revtex4}%
\usepackage{amsfonts}
\usepackage{amsmath}
\usepackage{amssymb}
\usepackage{graphicx}%
\begin{document}
\title {On the Classical Model for Microwave Induced Escape from a Josephson Washboard Potential}
\author{James A. Blackburn$^{1}$, Matteo Cirillo$^{2}$, and
Niels Gr{\o }nbech-Jensen$^{3}$}
\affiliation{$^{1)}$Department of Physics \& Computer Science, Wilfrid Laurier University,
Waterloo, Ontario N2L 3C5, Canada}
\affiliation{$^{2)}$Dipartimento di Fisica and MINAS-Lab, Universit\`{a} di Roma "Tor
Vergata", I-00173 Roma, Italy}
\affiliation{$^{3)}$Department of Applied Science, University of California, Davis,
California 95616}

\begin{abstract}
We revisit the interpretation of earlier low temperature experiments
on Josephson junctions under the influence of applied microwaves. It
was claimed that these experiments unambiguously established a
quantum phenomenology with discrete levels in shallow wells of the
washboard potential, and macroscopic quantum tunneling. We here
apply the previously developed classical theory to a direct
comparison with the original experimental observations, and we show
that the experimental data can be accurately represented
classically. Thus, our analysis questions the necessity of the
earlier quantum mechanical interpretation.

\end{abstract}
\maketitle

Since early days, Josephson junctions in superconducting circuits have been
modeled very successfully with the so-called Resistively and Capacitively
Shunted Junction (RCSJ) equivalent circuit \cite{VanDuzer}. In the situation
where a junction is biased with a fixed current, that equivalent circuit
yields an equation of motion for the junction phase that is analogous to a
fictitious particle confined to a "washboard" potential consisting of repeated
wells. In the presence of thermal noise, a Josephson junction biased just
below its critical current may be excited out of its position at the bottom of
a well and into a running state with an accompanying voltage. In 1981, a paper
appeared \cite{Voss} which reported experiments carried out on such a current
biased Josephson junction, but at much lower temperatures (down to $3mK$) than
previously. It was found that at high temperatures the transition rate
out of the metastable state was well described by thermal activation over the
barrier formed by the rim of the well, but that as $T\rightarrow0$, the
transition rate was dominated by quantum mechanical tunneling of the
macroscopic junction phase (MQT). The transition from classical to quantum
behavior was understood to occur in the neighborhood of a crossover
temperature \cite{Affleck}. \ A few years later \cite{Devoret}, Devoret {\it et al.}\
carried out another such experiment and again saw this evidence for MQT at the
lowest temperatures attained.

Beginning in 1985, Martinis, Devoret and Clarke \cite{Martinis,Martinis2}
applied these ideas to the interpretation of experiments which added
microwaves to the fixed current bias of an isolated Josephson junction. The
thinking was that because these experiments were carried out at low
temperatures ($18mK$ and $28mK$), only a quantum model need be considered.
Biasing the junction very close to its critical current makes the well very
shallow. Within the conceptual
framework of quantum phenomena, a well must contain discrete levels, and a
shallow well will contain only a few of them. The action of properly tuned
microwave radiation was then understood to populate the upper level from which
escape would occur via macroscopic quantum tunneling of the virtual particle
(i.e., the system would switch to the running state). The preferred
frequencies for enhanced escape observed in the experiment were thus taken as
signatures of the spacing of the discrete levels. The experimental data seemed
to confirm this model \cite{note} based on the quantized one-dimesional Josephson
washboard potential. This quantum picture was embraced by the
community of researchers interested in the possibility of using Josephson
based circuits as superconducting qubits.

However, no classical considerations were made in \cite{Martinis}, so the
argument as to an applicable model was in truth incomplete. In fact, as we
show here, the classical RCSJ model, which is the foundation for the above-mentioned
quantum phenomenology, with dc bias and applied microwaves
predicts the experimental observations to high accuracy.

The phase dynamics of a Josephson junction which is biased with both dc and ac
currents are governed by the simple classical expression \cite{VanDuzer}
\begin{equation}
\ddot{\varphi}+\alpha\dot{\varphi}+\sin\varphi=\eta+\eta_{d}\sin(\Omega
_{d}\tau) \label{Josephson} \; ,
\end{equation}
where the normalized dissipation is $\alpha=\hbar\omega_{0}/2eRI_{c}$, with
the Josephson zero-bias plasma frequency $\omega_{0}=\sqrt{2eI_{c}/\hbar C}.$
The junction RCSJ parameters are: critical current $I_{c}$, capacitance $C$,
resistance $R$. In this equation, the fixed and ac bias currents ($\eta$ and
$\eta_{d}$) have been normalized to $I_{c}$, the ac frequency has been
normalized to $\omega_{0}$, and time is in units of $\omega_{0}^{-1}$.
\ Equation (\ref{Josephson}) can be viewed as describing the classical motion
of a fictitious particle on a one dimensional surface%
\begin{equation}
U(\varphi)=-\eta\varphi-\cos\varphi\label{potential} \; ,
\end{equation}
with the potential expressed in units of the Josephson coupling
energy $E_{J}=\hbar I_{c}/2e$. This washboard potential has a
repeated series of wells which become shallower as the bias $\eta$
is increased. The natural frequency of small oscillations at the
bottom of each well changes with bias according to
$\omega_{p}=\omega_{0}(1-\eta^{2})^{\frac14 }$. The position of the
well minimum is $\varphi_{\min}=\sin^{-1}\eta$ and the rim is
located at $\varphi_{crit}=\pi-\varphi_{\min}$.

In the absence of ac excitation ($\eta_{d}=0$ in
Eq.(\ref{Josephson})), the particle would sit at the bottom of the
well. Sufficient additional thermal noise can activate escape over
the barrier $\Delta
U=2\left[\sqrt{1-\eta^{2}}-\eta\cos^{-1}\eta\right]$. Resonant
activation ($\eta_{d}>0$) provides an alternate mechanism for
escape, but the anharmonic shape of the well means that the ideal
condition is not simply to set the normalized forcing frequency
$\Omega_{d}$ to $(1-\eta^{2})^{\frac14 }$ ; generally speaking a
slightly lower value is optimum. It is also true that when the
forcing frequency is properly tuned to a specific well, there will
be some forcing amplitude above which escape will always occur. \
This means that just below such a drive threshold and when combined
with even very small thermal noise, stochastic activation will
occur, as is seen in all the experiments.

\begin{figure}
[pt]
\begin{center}
\includegraphics[
trim=0.0in 0.5in 0.0in 0.25in,
height=2.50000in,
width=3.00in
]%
{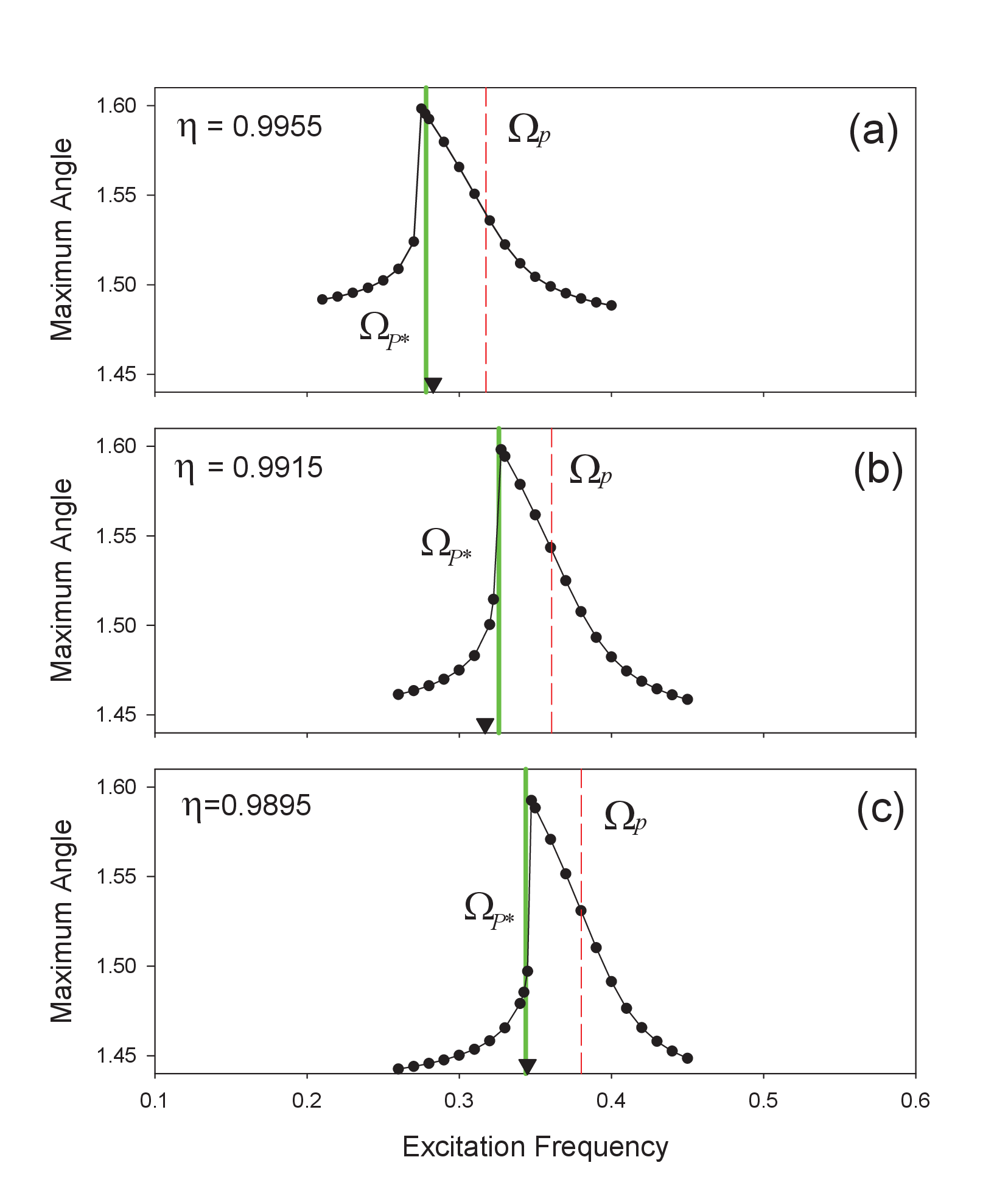}%
\caption{Plots of maximum amplitude of the junction phase oscillation on the
barrier side of the potential well as a function of the normalized
frequency of the applied microwave excitation for three different
normalized bias currents $\eta$ (numerical simulation). In each
case, the bias dependent normalized plasma frequency, determined
from the expression $\Omega _{P}=(1-\eta^{2})^{\frac14 }$, is shown
by a dashed vertical line. The anharmonic result $\Omega_{P\ast}$
from Eq.(\ref{exact}) is shown by a solid vertical line. The
normalized microwave frequencies from the experiments described in
\cite{Martinis} are marked by inverted black triangles.}%
\label{Fig.1}%
\end{center}
\end{figure}

A simple anharmonic analysis of the classical model can produce
explicit results that can be compared directly with experiments.
This was first done in Refs.~\cite{Jensen,anomalous}, where direct
comparisons between classical theory and experiments showed that
microwave-induced resonant switching can be understood classically.
The key results of the analysis come from introducing a
monochromatic ansatz $\varphi=\varphi_{0}+a\sin{\Omega_{p\ast}\tau}$
($\varphi_{0}$ is a constant and $a$ is the oscillation amplitude)
into Eq.(\ref{Josephson}) for $\alpha=\eta_{d}=0$. This yields the
effective relationships $J_{0}(a)\sin\varphi_{0}=\eta$ and
\begin{equation}
\Omega_{p\ast}^{2}=\frac{2J_{1}(a)}{a}\sqrt{1-\left(  \frac{\eta}{J_{0}
(a)}\right)  ^{2}}\; , \label{exact}%
\end{equation}
where $J_{n}$ is the Bessel function of $n$th order and first kind. Since the
experiments are conducted such that high escape probability exists near the
sought-after resonance while vanishing escape probability should exist when
the system is off-resonance, the strength of the microwave field should be
such that the resulting oscillation amplitude reaches the inflection point of
the effective potential; i.e.,
\begin{equation}
a=\frac{\pi}{2}-\sin^{-1}\frac{\eta}{J_{0}(a)}\; \label{parameter} \; .
\end{equation}
Oscillating with this amplitude will result in a good statistical
probability that the junction switches to the non-zero voltage
state within some allowed time. A useful explicit second order
approximation in $a$ for
$\eta\rightarrow1$ to the desired amplitude is given by
$a^{2}=\frac{4}{3}(1-\eta)$. Inserting this result into
Eq.(\ref{exact}) provides an explicit relationship between the
anharmonic classical resonance and the bias current for direct
comparisons with reported experiments in the limit of validity for
Eq.(\ref{parameter}).
\begin{table}[htdp]
\begin{center}
\includegraphics[
trim=0.0in 0.5in 0.0in 0.0in,
height=0.75000in,
width=3.00in
]%
{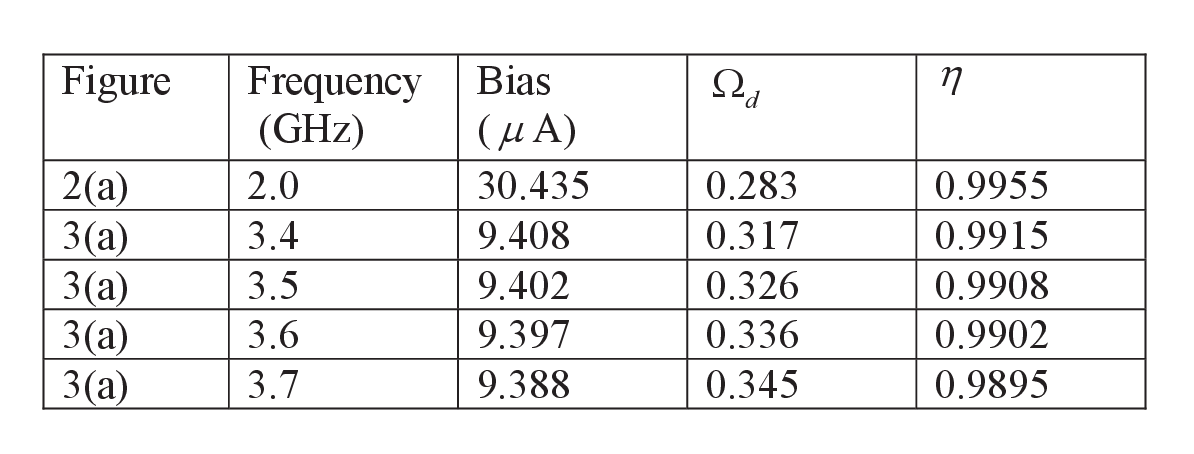}%
\end{center}
\caption{Summary of experimental data for the peaks in the escape rates as
    reported in Martinis {\it et al.}\ [5].}
\label{Table1}
\end{table}%
The protocol adopted in the experiments \cite{Martinis, Jensen} was
to fix the microwave frequency and then observe the escape rate
while scanning the bias current. Our attention here is on the
results originally presented in \cite{Martinis}, most particularly
the data in their Figs. 2 and 3. Two different physical samples were
used - the parameters for Fig.2 were given as $I_{c}=30.572 \mu A$,
$C=47.0 pF $, yielding a plasma frequency
$\omega_{0}=4.443\times10^{10} s ^{-1}$, whereas for Fig. 3 they
were $I_{c}=9.489 \mu A$ , $C=6.35 pF $, giving a plasma frequency
$\omega_{0}=6.734\times10^{10} s ^{-1}$. From blowups of these
figures, it was possible to identify the peak positions, that is the
bias current at maximum escape rate for each stated microwave
frequency -- these are given in the second and third columns of
Table I. Normalized values $\Omega_{d}$ and $\eta$ are listed in the
fourth and fifth columns. The most prominent peak in Fig. 2 and all
four peaks in Fig. 3 are included.
\begin{figure}
[pt]
\begin{center}
\includegraphics[
trim=0.0in 0.5in 0.0in 0.0in,
height=1.750000in,
width=3.00in
]%
{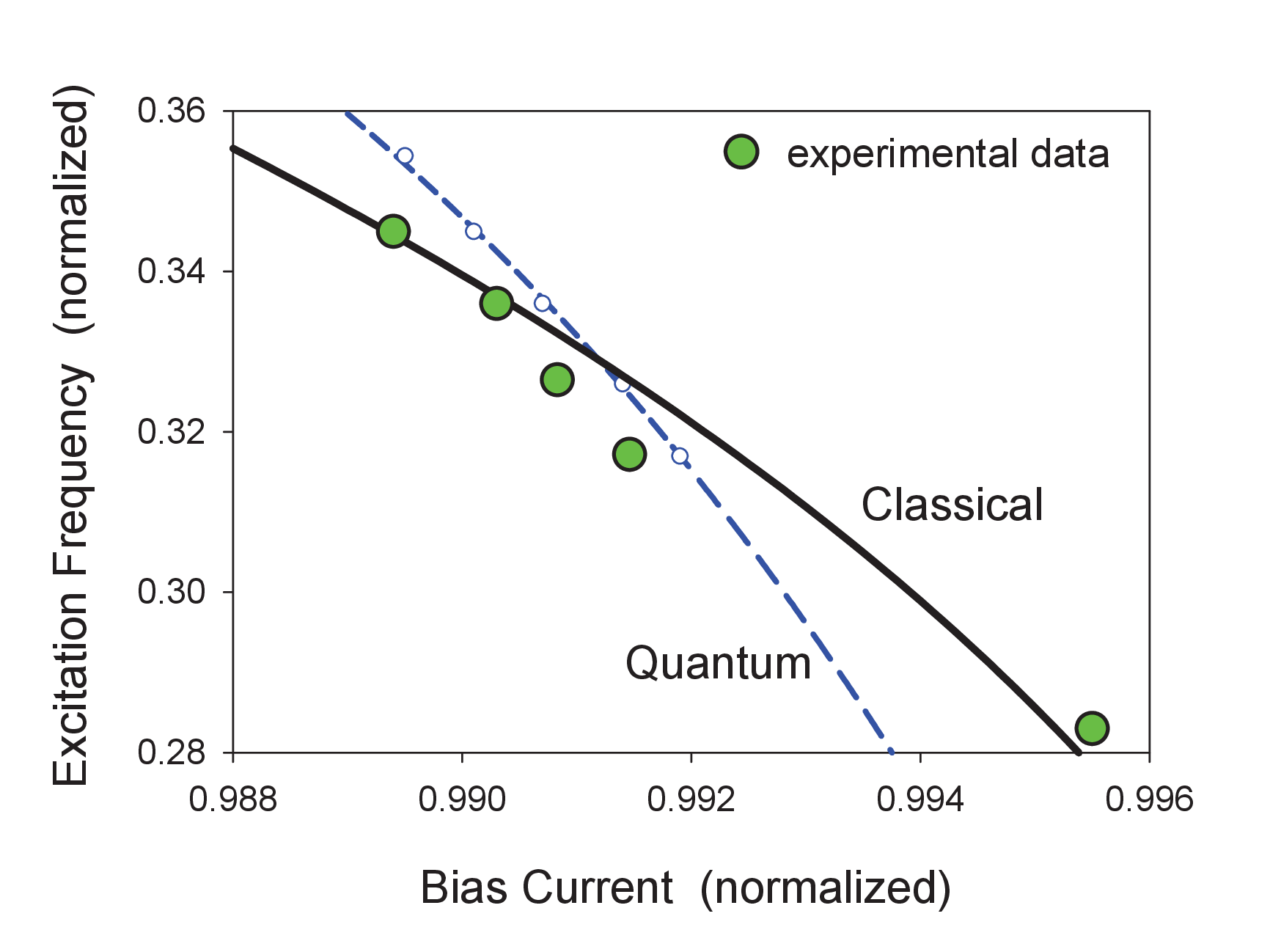}%
\caption{A comparison of the classical and quantum models for the escape to a
finite voltage state of a low temperature Josephson junction biased very close
to its critical current and with applied microwaves. The 5 experimental data
points are from Martinis {\it et al.}\ \cite{Martinis}, as is their predicted quantum
result (dashed line). The anharmonic analysis of the classical model yields
the solid line.}%
\label{Fig.2}%
\end{center}
\end{figure}
It should be noted that because the precise shapes of the potential
wells are completely determined by the bias current $\eta$, it is
somewhat more physically meaningful, in contrast to the experimental
protocol mentioned above, to fix the bias current and vary the
excitation frequency. This approach reveals the optimum excitation
frequency for any given well (bias). Such a strategy was followed in
carrying out numerical solutions of Eq.(\ref{Josephson}). In these
simulations, the value of the dissipation constant $\alpha$ was
chosen to be $0.00845$, otherwise there were no adjustable
parameters. For any selected bias $\eta$ matching one of the
experimental values, a series of runs was carried out for specific
frequencies $\Omega_{d}$. The excitation amplitude $\eta_{d}$ was
ramped up from zero to its final value (typically in the range
$0.0007$ to $0.001$) over an interval of $600$ plasma periods; this
was done to avoid spurious start up effects. An additional $400$
plasma periods were then recorded. In each run the resulting data
set from the numerical solution $\varphi(\tau)$ was examined for the
largest repetitive maximum angle reached. These maximum angles are
plotted in the three panels of Fig.1. They correspond to the first,
second and fifth rows of Table I. Computations corresponding to the
other bias values yield the same general behavior. The obvious
asymmetries of the responses to ac excitation are a result of the
anharmonicity of the wells. The inverted filled triangles mark the
frequencies of the experimentally determined peaks in the escape
rates, as reported in \cite{Martinis}. These frequencies coincide
almost exactly with the location of the sudden onset of large
induced phase oscillations in the numerical solutions of
Eq.(\ref{Josephson}). These plots also reveal that good agreement
should not be expected with the normalized linear plasma frequency
$\Omega_{p}$ because it is always an over-estimate of the true
optimum excitation for which a better expression is given by
Eqs.~(\ref{exact}) and (\ref{parameter}).

Figure 2 presents a direct comparison of the classical and quantum models with
respect to the experimental data reported in \cite{Martinis}. The left-most
four points come from the four peaks presented in Fig. 3(a) of \cite{Martinis}%
; the fifth data point, on the far right, comes from the position of
the right-most peak in Fig. 2(a) of \cite{Martinis}. The dashed line
labeled \textquotedblleft Quantum\textquotedblright\ is obtained
from the predicted energy level spacing $E_{0\rightarrow1}$ shown as
a solid curve in Fig. 3(b) of \cite{Martinis}; the solid curve
labeled \textquotedblleft Classical\textquotedblright\ is generated
from Eqs.~(\ref{exact}) and (\ref{parameter}).

As has been demonstrated, in the absence of noise, optimum ac
excitation will induce a maximum excursion in the phase of the
Josephson junction. These largest excursions take the system closest
to the critical value $\varphi_{crit}$, at which escape would occur.
Our basic hypothesis is simple: for a slightly subcritical
excitation, additional noise will stochastically induce escapes, as
was demonstrated in \cite{anomalous}. Therefore the experimental
observations of peaks in the escape rates at finite temperature are
fully consistent with the classical model embodied in Eq.
(\ref{Josephson}). \ Agreement between predictions of the classical
model with experimental values for the dc bias ($\eta$) and
microwave frequency ($\Omega_{d}$) at these peaks is very accurate.
Moreover, the classical model maintains its accuracy over two
distinct experimental samples having quite different critical
currents and capacitances.
\begin{figure}
[pt]
\begin{center}
\includegraphics[
trim=0.0in 0.5in 0.0in 0.0in,
height=2.50000in,
width=3.00in
]%
{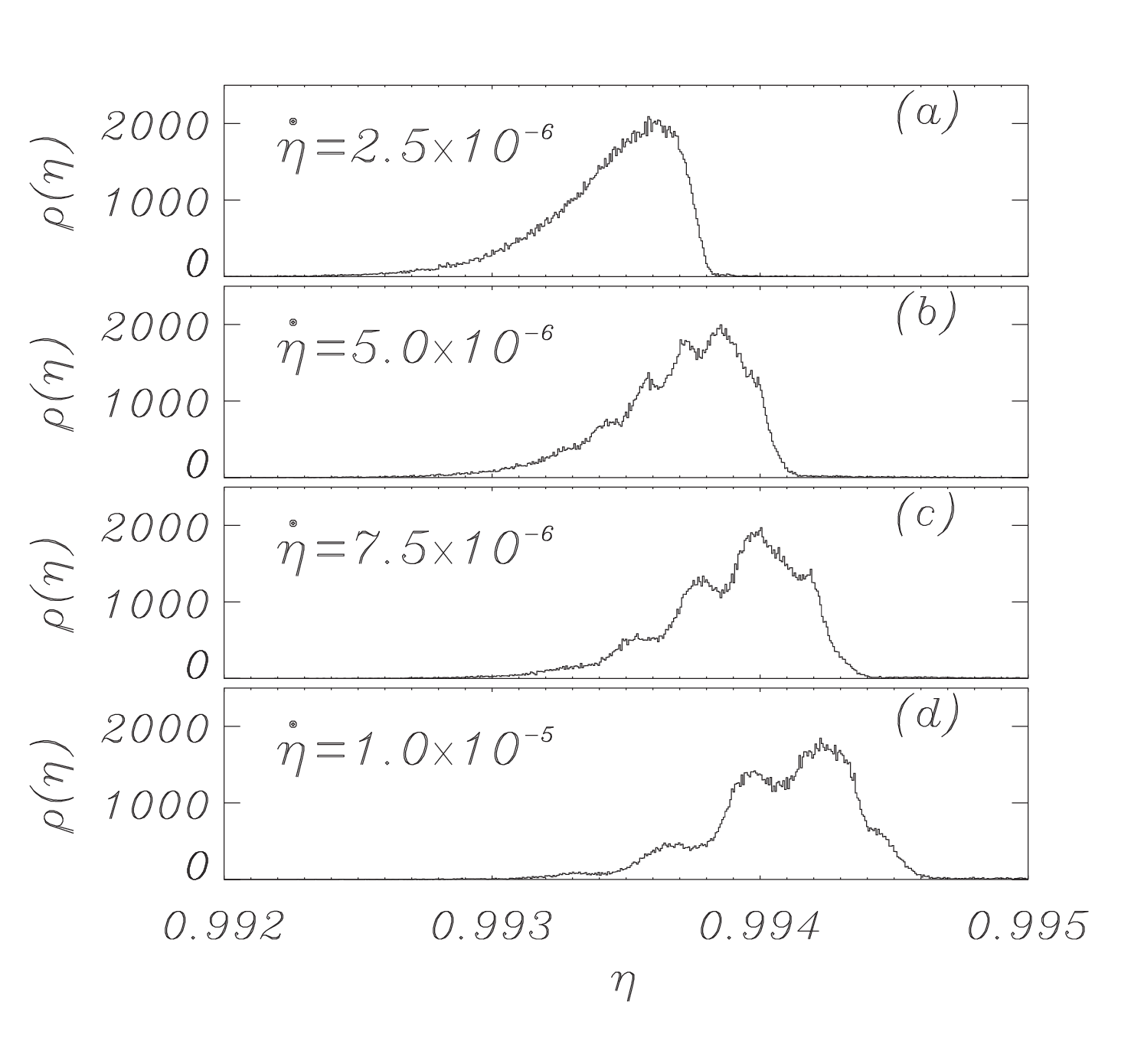}%
\caption{Switching distributions for swept bias $\eta$. For these simulations, we chose
    $k_BT/E_J=3.8418\times10^{-5}$ and $\Omega_d=0.282809$ to match the 2 GHz frequency in
    Fig. 2 of \cite{Martinis}. Each distribution is generated from
    100,000 switching events, and each simulation is initiated at $\eta=0$.
    The applied sweep rate, microwave amplitude and damping were not reported
    in \cite{Martinis}; we have chosen $\alpha=10^{-4}$ and $\eta_d=0.0019$. The
    applied sweep rate $\dot{\eta}$ is reported on each panel.}%
\label{Fig.1}%
\end{center}
\end{figure}
One curious observation made in \cite{Martinis} is the appearance of
multiple resonant escape peaks seen in their Figure 2. These were
interpreted as evidence for the higher quantum transitions
$1\rightarrow2$, $2\rightarrow3$, and so on. However, no signatures
of these were seen in the comparable escape probability plots shown
in their Figure 3. We further notice that such a family of peaks was
also not observed in \cite{Ustinov}. Our simulations of the
classical model show that closely spaced multiple resonant escape
peaks can be observed under certain conditions due to the
nonlinearity that provides the possibility for phase-locking of an
oscillator at many harmonics (sub, super, and fractional). We
mention in passing that this has also been shown \cite{Jensen} to
explain the so-called multi-photon absorption peaks discussed in
\cite{Ustinov}. One could imagine that the multiple resonances seen
in Figure 2 of \cite{Martinis} could be due to some fractional
resonance, but we see no evidence for this in the classical model
for the given parameters. Instead, we have noticed very similar
families of resonant peaks appearing when the \textit{bias current
sweep rate} is elevated. This important parameter is not given in
\cite{Martinis}, but typical values would correspond to a normalized
sweep rate in the interval between $\sim10^{-8}$ and $\sim10^{-5}$.
It is in the upper end of this range (faster sweep rates) that we
encounter the appearance of the multiple peaks similar to those
observed in \cite{Martinis}. Figure 3 shows the appearance of such a
switching peak family as a function of the bias sweep rate
$\dot{\eta}$ for a fixed microwave amplitude . All other parameters
are chosen according to the experiments leading to figure 2 in
\cite{Martinis}. Thus we included in the right hand side of
Eq.(\ref{Josephson}) a stochastic current noise term
\cite{Jensen,anomalous} corresponding to the thermodynamic
temperature of $28mK$; each distribution in Fig. 3 is the result of
100,000 switching events after initiating the system at $\eta=0$.

We clearly observe in Fig. 3 the emergence of multiple peaks, and we
further observe that the spacing between the peaks increases with
the sweep rate. We have investigated the details of the resonance
near the bias values of the multiple peaks, and we have found only a
single resonance, which is well described by Eqs.~(\ref{exact}) and
(\ref{parameter}), and this is consistent with the fact that only a
single resonant peak is found for small sweep rates. Thus, the
multiplicity of apparent resonant peaks must be related to an
interference between the microwave oscillation and the
non-negligible variation in $\eta$ for high sweep rates. In this
context it is worthwhile recalling that a fast bias current sweep
rate has previously been shown to produce multi-peak switching
distributions in simulations of long Josephson junctions
\cite{Jensen04}, and it has been shown experimentally that fast
sweep rates can produce multiple peaks in the conditional switching
distribution \cite{Silvestrini} even without the application of
microwaves.

To conclude, we have established that the postulated quantum ingredients of
discrete levels and macroscopic quantum tunneling are superfluous in
explaining the overall results of experiments on resonant escape in current biased Josephson
junctions at low temperature and with applied microwaves; the well-known
classical model has been shown to yield excellent agreement with the reported
experimental data. This agreement is consistent with the ones previously
reported in Refs. \cite{Jensen,anomalous}, and is quite similar to our recent
examination of claims of quantum entanglement in coupled Josephson qubits
\cite{Blackburn}, and to our earlier work \cite{Jensen1} showing that the
classical model accounts for reports of, e.g., Rabi oscillations in large Josephson
junction qubits \cite{Martinis3}.

The often repeated claim \cite{Clarke} that the phase variable of
Josephson junctions in these superconducting systems at these
temperatures has been \textquotedblleft
unequivocally\textquotedblright\ demonstrated to be exclusively a
quantum variable is not supported by the present evidence. What is
lacking are experimental observations of phenomena that cannot be
encompassed within a classical picture and thus require an
exclusively quantum model. The classical RCSJ model seems adequate
in seeking an overall understanding of these types of experiment
where microwaves are applied to Josephson systems.

\begin{acknowledgments}
This work was supported in part (JAB) by a grant from the Natural Sciences and
Engineering Research Council of Canada.
\end{acknowledgments}

\end{document}